%% file: main.tex
\documentclass[conference,10pt]{IEEEtran}
\usepackage[final]{graphicx}
\usepackage{multicol}
\graphicspath{{Images//}} 
\usepackage{datetime}
\usepackage{hyperref}
\usepackage{psfrag}
\usepackage{epstopdf}
\usepackage{amsfonts}
\usepackage{mathrsfs}
\usepackage{pifont}
\usepackage{verbatim}
\usepackage{upgreek}
\usepackage{color}
\usepackage[usenames,dvipsnames]{xcolor}
\usepackage{algorithm}
\usepackage{mdwmath}
\usepackage{mdwtab}
\usepackage{amssymb,amsmath}
\usepackage{amsmath}
\usepackage{amsthm}
\usepackage{epstopdf}
\usepackage{cite}
\usepackage{tikz}
\usepackage{scalefnt}
\usepackage{subfigure}
\usepackage{MnSymbol}
\usepackage{setspace}
\usepackage{mathtools}
\usepackage{psfrag}

\usepackage[font=scriptsize]{caption}
\usepackage{algpseudocode}

\UseRawInputEncoding

\algrenewcommand\algorithmicforall{\textbf{foreach}}
\algrenewcommand\algorithmicindent{.8em}

\label{newcommand}

\label{English Chars}

\label{Roman Chars}

\label{Theorems}

\theoremstyle{remark}

\begin{document}
\label{title}
\title{A review of the security role of\\
ISP mandated ONUs and ONTs in GPONs}

\author{%
\IEEEauthorblockN{Max Franke}
\IEEEauthorblockA{\textit{INET} \\
\textit{Technische Universit{\"a}t Berlin}\\
Berlin, Germany}
\and
\IEEEauthorblockN{Sebastian Neef}
\IEEEauthorblockA{\textit{SecT} \\
\textit{Technische Universit{\"a}t Berlin}\\
Berlin, Germany}
}

\maketitle

\begin{abstract}
\input{sections/0-Abstract}
\end{abstract}

\input{sections/1-Introduction}
\input{sections/2-Background}

\input{sections/3-Analysis}
\input{sections/4-Results}

\input{sections/5-Conclusion}

%


\ifCLASSOPTIONcaptionsoff
\newpage
\fi
\bibliographystyle{ieeetr}
\bibliography{ref}

\end{document}

%% file: sections/0-Abstract.tex
Home fiber connections are largely realized by using passive optical networks, in their most common form today relying on the GPON standard. 
Among other things, this standard specifies how the first node inside of customers' homes, the so called ONU or ONT, has to behave, and which security features have to be supported. 
Currently, customers in some European countries, including Germany, have freedom of choice between using terminal equipment provided by the ISP or a self-selected open market device.
We analyze the security implications resulting from this freedom of choice and whether or not ISP-mandated hardware would increase the security of the GPON. 
Our review reveals that there are no differences between an ISP-mandated ONU/ONT and a standard conforming subscriber-selected ONU/ONT that would justify the security based recommendation of an ISP-mandated ONU/ONT.

%% file: sections/1-Introduction.tex
\section{Introduction} 
With the ever-increasing demand for higher bandwidth and the consequent expansion of fiber networks by Internet Service Providers (ISPs), like Fiber-to-the-Home (FTTH), reaching more and more customers, it becomes crucial to focus on the technologies and devices that facilitate this progress. 
The prevailing standard for such connections is the gigabit passive optical network (GPON), which was initially standardized by the ITU-T in 2003. 
The market penetration of FTTH utilizing GPON can reach as high as 97\% in certain markets, such as the UAE. 
However, the adoption has been comparatively low in many European countries, with Germany for instance at around 8\% in June 2022\cite{OECD_2022}, but targeted to increase to 50\% by 2025 \cite{Bundesregierung_2023}.

The GPON standards encompass numerous security measures and considerations for both the optical line terminal (OLT), responsible for terminating the connection at the provider's end, and the optical network unit (ONU, also "optical network termination"-- ONT) at the customer's end. 
These standards are mandatory for all devices, regardless of whether they are provided by ISPs or third-party equipment vendors. 
The permissibility of operating third-party devices depends on the legislation in each country. 
In Germany, all network subscribers are granted by law\footnote{See e.g. https://www.bmwk.de/Redaktion/DE/Artikel/Digitale-Welt/freie-routerwahl.html 
} the freedom to choose their own terminal device. This has been applied to fixed and mobile access networks in general, including DSL, DOCSIS, PON fiber, and 4G/5G Fixed Wireless Access.

In most cases, the PON terminal equipment combines the functionalities of both a router and a modem, resulting e.g. in space and power savings at the users' premises.
When such an integrated device is used as the terminal device, it serves the function of an ONU/ONT for the subscriber in the fiber network.
It is also possible to split the functionalities and use two separate devices, a modem and a router.
In this case, the modem serves the function of an ONU/ONT. 

As such, a vendor of terminal equipment approached us to investigate the question whether an ISP-mandated ONU/ONT offers any security advantages compared to a standard conforming open market ONU/ONT. Our review considers any security differences affecting the subscriber itself, the ISP's network, and the security of other subscribers on the same PON.   

%% file: sections/2-Background.tex
\section{Background}

\subsection{ITU-T Standards}
The GPON standard was initially standardized in 2003 as G.984. It consists of a multitude of documents \cite{itu-t_g9841_nodate,itu-t_g9842_nodate,itu-t_g9843_nodate,itu-t_g9844_nodate,itu-t_g9845_nodate,itu-t_g9846_nodate,itu-t_g9847_nodate}, each covering specific aspects of GPON.
The details for the information exchange between an OLT and an ONU are described in G.988, the ONU management and control interface (OMCI) specification \cite{itu-t_g988_nodate}. 
Closely related is ITU-T G.9807.1 (XGS-PON), which increases the bandwidth of GPONs to up to 10 GBit up- and downstream \cite{itu-t_g98071_nodate}. 
However, this work does not consider XGS-PON.

\subsection{Definition of terms: ONU and ONT}
The definitions of ONU and ONT in \cite{itu-t_g9841_nodate} are as follows: 

\begin{quote}
    \textbf{optical network termination (ONT)}: A single subscriber device that terminates any one of the distributed (leaf) endpoints of an ODN, implements a PON protocol, and adapts PON PDUs to subscriber service interfaces. An ONT is a special case of an ONU.
\end{quote}

and 

\begin{quote}
    \textbf{optical network unit (ONU)}: A generic term denoting a device that terminates any one of the distributed (leaf) endpoints of an ODN, implements a PON protocol, and adapts PON PDUs to subscriber service interfaces. In some contexts, an ONU implies a multiple-subscriber device.
\end{quote}

\cite{itu-t_g9841_nodate} makes no further differentiations when it comes to ONU and ONT and often uses them either interchangeably or both at the same time. This is also true in the security section:
\begin{quote}
    \textbf{Security}
Due to the multicast nature of the PON, GPON needs a security mechanism adapting the following requirements:
\begin{itemize}
    \item  To prevent other users from easily decoding the downstream data.
    \item To prevent other users from masquerading as another ONU/ONT or user.
    \item To allow cost-effective implementation.
\end{itemize}
\end{quote}

As the ONT is just a special case of ONU, none of the other specifications make any further differentiation when it comes to security considerations. In fact the specification in which most of the security related aspects are included, \cite{itu-t_g988_nodate}, makes no references to ONTs at all.

\subsection{Related Work}
Twenty years after the initial standardization, \cite{horvath_security_2015} highlighted security problems in and proposed enhancements for GPON.

From a practical perspective, there have been several reports of security issues in GPON setups or capable devices \cite{kim_gpon_nodate,newman_critical_nodate,martingale_is_2021}. 

There have also been works that review security flaws in other telecommunication standards. E.g., \cite{camenisch_case_2011} demonstrated an attack on DOCSIS over ten years ago. 
With GPON, however, a discussion arose about the point of network termination.
One could argue that the network termination happens at the fiber in the wall, while others might argue that the GPON modem is part of the network and thus the network termination follows after it.  

Network security can be one of the deciding factors in this debate. It is therefore necessary to establish whether the use of an ISP-mandated device instead of a subscriber-selected standard conforming device is crucial to ensure the security of the PON.

%% file: sections/3-Analysis.tex
\section{Research Questions and Methodology}

\begin{figure}[h]
    \centering
    \includegraphics[width=0.5\linewidth]{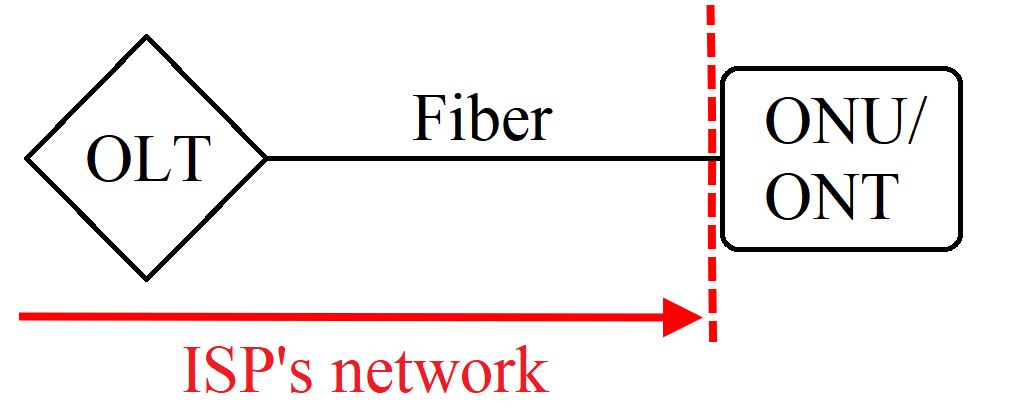}
    \caption{Scenario 1 with subscriber-selected ONU/ONT. The ISP's network includes the fiber, but excludes the subscriber's ONU/ONT. }
    \label{fig:scenario1}
\end{figure}

\begin{figure}[h]
    \centering
    \includegraphics[width=0.5\linewidth]{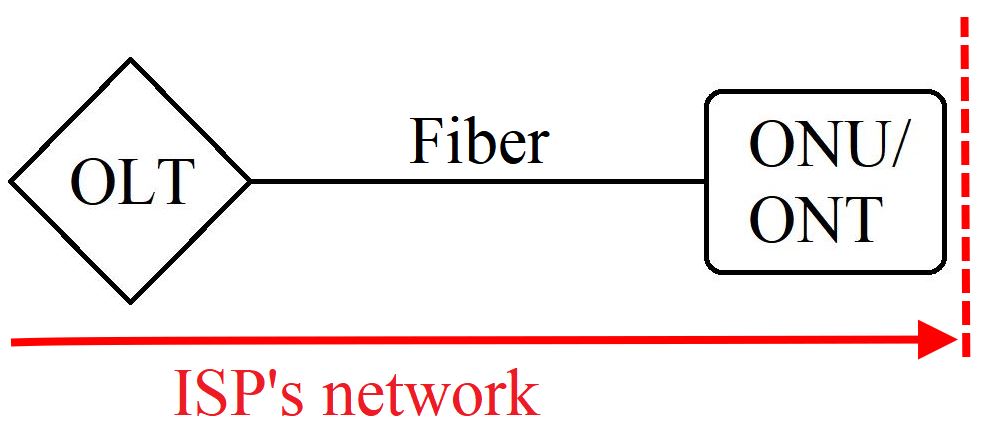}
    \caption{Scenario 2 with ISP-mandated ONU/ONT. The ISP's network includes the fiber as well as the subscriber's ONU/ONT. }
    \label{fig:scenario2}
\end{figure}

We evaluate the security of GPON concerning the two scenarios illustrated in Figure \ref{fig:scenario1}  and Figure \ref{fig:scenario2} . 

In scenario 1, the ISP's network ends before the ONU/ONT, and the ONU/ONT can be selected by the subscriber. 

In scenario 2, the ISP's network ends after the ONU/ONT, and the ONU/ONT is mandated by the ISP.

\subsection{Research questions}
Our review considers the following research questions:
\begin{enumerate}
    \item[RQ1] Assuming ITU-T conforming ONUs/ONTs, is the security level of a PON in scenario 1 (subscriber-selected ONU/ONT) at least as high as in scenario 2 (ISP-mandated ONU/ONT) according to the ITU-T standards?
    \item[RQ2] According to the ITU-T standards, can an ISP's PON network equipment enforce a minimum level of security, so that an ONU/ONT is not unilaterally able to reduce this level?
    \item[RQ3] Are ONUs/ONTs, and thus the subscriber, assigned the task of maintaining network security? 
    \item[RQ4] Does the ONU/ONT have to be part of the ISP's network to guarantee security?
    \item[RQ5] Are all the security features the standard offers required to be used? 
\end{enumerate}

\subsection{Threat model}
The vendor suggested the following boundary conditions as scope for the research threat model, matching the strong focus on the ITU-T standards:
\begin{itemize}
    \item All devices are standard-compliant and properly integrated into the PON.
    \item Scenarios of malicious attackers with physical access to the fiber or to the intermediary network equipment, i.e. the passive optical components in the basement or on the street, are excluded. Such scenarios do not imply any differences between ISP-mandated and subscriber-selected devices and are therefore not relevant for our review. 
\end{itemize}

\subsection{Methodology}
After a literature review of existing academic and non-academic work, our methodology consists of going through the relevant parts of the GPON ITU-T standards with a focus on aspects relevant to our security considerations. 
Additionally, the vendor provided us a trace of a GPON connection initialization between an ONU/ONT and OLT.

%% file: sections/4-Results.tex
\section{Results}
With the given threat model and methodology, our analysis results in the following answers to our research questions: 

\subsection{RQ1}
Yes, the security levels are identical for scenario 1 and scenario 2, assuming ITU-T conforming devices. Our review does not show any differences in the standards regarding the two scenarios. The standards do not differentiate between both scenarios, neither in general nor with respect to security. 
Even the term "Network Termination" (NT) that is used within the standards does not imply an association of a terminal device to an ISP's network. The ITU itself states \cite{itu-t_recgsup50}:
\begin{quote}
    The NT term is used for generic network
termination for various services. For some services it could be part of the access network, and for
others not. The inclusion of the NT in the access network and vice versa does not necessarily imply
the ownership.

[...]

The reference configurations in this clause show abstract functional groupings, which may or may
not correspond to real devices. Real devices may comprise one abstract functional grouping, more
than one abstract functional grouping or a portion of an abstract functional grouping.
\end{quote}

In practice, all devices in a PON will need to support the required security features and properties, e.g. cryptographic algorithms. These can be implementation specific and vary from PON to PON. Nonetheless, it does neither weaken nor impact scenario 1, as conforming subscriber-selected devices will offer the required functionality and only need to provide a suitable configuration interface.

Therefore we do not see any justification from the standards to propose a security advantage arising from ISP-mandated ONUs/ONTs over conforming subscriber-selected ONUs/ONTs.

\subsection{RQ2}
Yes, an ISP's PON network equipment can enforce a minimum level of security, so that an ONU/ONT is not unilaterally able to reduce this level. The desired security properties are exchanged and negotiated between the OLT and ONU/ONT.  In practice, an ONU/ONT will not be able to reach the state "O5-Operational" if the OLT does not approve -- e.g., because the OLT's requests to choose a session key are not answered properly by the ONU/ONT.

Thus, it is left to the implementation of the OLT to require and enforce acceptable security properties.

\subsection{RQ3}
In GPON, the downstream is received by all ONUs/ONTs sharing the same fiber to an OLT. The standards recommend the use of encryption to ensure only the legitimate ONU/ONT can decipher its own downstream traffic. In this regard, the ONUs/ONTs are a contributing factor to the security for its own downstream, but it has no influence on the security of the network or other subscribers of the PON. 

Regardless of the considered scenario, the subscriber has to configure the ISP-mandated or subscriber-selected ONU/ONT with specific credentials. However, this is not a new task or role, as the subscriber already needs to provide the ISP with ONU/ONT-specific information in current GPON deployments, or needs to configure the router with correct credentials in existing DSL deployments. Obviously, it is in a subscriber's best interest to correctly configure the device, as misconfigured devices might not provide the desired internet connectivity. No security differences between scenario 1 and 2 can be deduced here.

\subsection{RQ4}
As the location of the ONU/ONT is inside subscribers' homes and easily accessible by them, they can not be required to be part of the ISP's network to guarantee security. A malicious actor could simply remove the modem and attach any device to the fiber, thus breaking security. As such, security has to be ensured and enforced by the OLT.  
An ONU/ONT with a non-conforming implementation or security flaws in the implementation would be far from the worst-case that could happen in such scenarios. 

Thus, the position of a standard conforming ONU/ONT -- inside or outside the ISP's network -- has no effect on the rest of the network.
A standard conforming implementation, including all optional security controls (see RQ5), provides the same security guarantees to the network, disregarding the origin of the device. 

\subsection{RQ5}
The standard \cite{itu-t_g988_nodate} includes optional so called ''Enhanced security control'' that can be negotiated. Among those additional security features is a Pre-Shared Key (PSK) that would be used in e.g. different hash functions. The example trace of a GPON connection initialization did not show the use of a PSK. From discussions with the equipment vendor, these enhanced security features are typically not activated or used by actual PON network operators.

There are several other optional security features the standard allows for. As such, to further enhance the security of the GPON, ISPs should work on implementing and utilizing these additional capabilities -- and third-party vendors support them.
Additionally, newer standards like XGS-PON include features like upstream encryption that further reduce the role the ONU/ONT takes in guaranteeing security for the network. 

\subsection{Limitations}
The defined threat model imposed strong limits on possible attack scenarios against a PON or its users, which may arise from (physical) access to the fiber, including non-standard behavior or messages on the protocol level.

In reality, physical access can hardly be prevented. For example, in a FTTH setup, the fiber terminates at \emph{home} in a ONU/ONT. For FTTB/C and FTTCab, reports of optical network equipment present in customer or publicly accessible places (i.e. basements of multi-story building) exist \cite{kim_gpon_nodate}.
As a consequence, such attack scenarios against PONs need to be considered and mitigated at the OLT level, as the resulting security risks are not due to a specific type of ONU/ONT being used.

%% file: sections/5-Conclusion.tex
\section{Conclusion} 
The analysis of the security role of ONUs/ONTs in GPON shows that the standards do not differentiate between ISP-mandated and subscriber-selected devices. Differences between both scenarios with regards to security or the security capabilitites can not be deduced. 

Assuming standard conforming devices, the security of PONs is not influenced by the use of ISP-mandated or subscriber-selected devices, as long as the ISPs provide the necessary configuration information (i.e. credentials, PSKs, etc.) and third-party devices provide a suitable configuration interface.
The minimal requirements for the networks will need to be enforced by the OLT during the exchange of capabilities. The OLT can even ensure and enforce a certain security level for the individual subscriber. There is no difference in security for the network if the ONU/ONT is part of the ISP network or not.

The individual implementation of the standard on any device is specific to this device. It is therefore not possible to make any general statements on security differences between ISP-mandated and subscriber-selected ONU/ONTs. And even if there are potential security vulnerabilities within a device-specific implementation, this cannot compromise the security of other subscribers of the PON.

In reality, malicious attackers usually aim to threaten the security of the network itself. Security threats against GPONs are therefore far more likely to arise from rogue, non-standard conforming devices or physical access to the fiber. Such malicious attack scenarios, which were discussed in previous work, are agnostic to the chosen device and the association to the ISP's network.